\documentclass[twocolumn,showpacs,preprintnumbers,amsmath,amssymb]{revtex4}

\usepackage{graphicx}
\usepackage{dcolumn}
\usepackage{bm}

\begin{document}


\title{Sintering behaviour of two roughened crystals just after contact}

\author{Robert S. Farr}
 \affiliation{Unilever R\&D, Olivier van Noortlaan 120, AT3133, Vlaardingen, The Netherlands}
 \email{robert.farr@unilever.com}
\author{Martin J. Izzard}
 \affiliation{Unilever R\&D, Colworth House, Sharnbrook, Bedford, England,
MK441LQ}

\date{\today}

\begin{abstract} We consider two spherical, roughened crystals 
with approximately
isotropic surface free energy which are brought into contact and begin
to sinter. We argue that the geometry 
immediately post-contact is two dimensional and Cartesian and can
be approximated by the evolution of a slot-shaped cavity.
On this basis, we construct
travelling wave solutions for the crystal shape in the limits of
bulk diffusion limited and surface diffusion limited kinetics. These
solutions are then used to calculate novel scalings for the neck size 
as a function of time $t$ after contact: We predict that neck size is 
proportional to $t^{1/4}$ for the bulk diffusion limited case and (following
a single pinch-off event) approximately proportional to $t^{1/3}$ for
the surface diffusion limited case.
\end{abstract}

\pacs{81.10.Aj}

\maketitle

\section{Introduction}
A crystal in equilibrium against its vapor or a melt eventually reaches a
shape which can be obtained by the Wulff construction \cite{Wulff}:
Let the free energy per unit area of a plane surface of the crystal which is
perpendicular to a unit vector ${\rm\bf \hat{n}}$ be given by 
$\gamma ({\rm\bf \hat{n}})$. We define the points
${\rm\bf p}({\rm\bf \hat{n}})\propto {\rm\bf \hat{n}}\gamma ({\rm\bf \hat{n}})$
for some fixed constant of proportionality which eventually sets the size of the
crystal. Next we construct a plane through each ${\rm\bf p}({\rm\bf \hat{n}})$, 
perpendicular to ${\rm\bf \hat{n}}$. The Wulff shape of the crystal 
is the inner envelope of all such planes.

From the terrace-ledge-kink (TLK) \cite{Kossel,Stranski} model
we expect the surface free energy of a crystal to have cusp-like
minima near to the crystallographic symmetry directions: For a plane oriented
at a small angle $\theta_v$ to a crystallographic axis (i.e. a `vicinal
plane') and at absolute temperature $T$, the surface free
energy per unit area behaves like \cite{Williams}
\begin{equation}
\gamma(\theta_v,T)\approx\gamma_0(T) + \frac{\sigma(T)}{d_m^2}\left|
\tan\theta_v\right|,
\end{equation}
where $\gamma_0$ is the free energy per unit area of a molecularly flat 
surface with $\theta_v=0$, while $\sigma$ 
is the ledge energy per molecule and $d_m$ is a molecular distance.

Using the
Wulff construction, this leads to the emergence of crystal facets
at equilibrium, in which a macroscopic portion of the crystal surface is
molecularly flat, save for isolated single molecule islands and
surface vacancies \cite{Burbon}.

However, at a particular temperature (the `equilibrium roughening' 
or `surface melting' temperature), the free energy barrier to formation of new
molecular islands on the facet vanishes;
the facets become rough on a molecular scale, so that the
relevant cusp in the surface free energy disappears along with the facet,
to be replaced by a macroscopically smooth, rounded crystal surface
\cite{Burbon,Pavlovska}. This roughening transition is different for
different crystallographic symmetry directions; for example three
different temperatures have been observed for Helium crystals
\cite{Wolf}, while for ice, the basal facet persists up until melting,
while prism planes roughen at $-2^\circ$C when against vapor \cite{Elbaum} or 
$-16^\circ$C when against water under pressure \cite{Manuyama}.
If the crystal surface is not at equilibrium, but instead growing, this can
move the roughening transition to lower temperatures 
\cite{HalpinHealy,Williamson}.

In this investigation, we start from two assumptions: 
First, we assume that all the 
relevant crystallographic planes are roughened (there may however be other 
directions not involved in the analysis described below, which are faceted).
Second, we make the approximation that the roughened portion of a crystal 
surface has a fairly isotropic surface free energy per unit area $\gamma$
(a typical approximation for phase-field studies 
of dendritic growth \cite{Karma} or solvability theory \cite{Barbieri}).

When a roughened crystal does not have its Wulff shape, then the surface 
free energy will not be equal everywhere.  The chemical potential $\mu$ of a 
molecule at a curved surface differs from the value $\mu_0$ it would 
have for a flat surface by the well known Gibbs-Thomson equation, which 
in linearized form gives
\begin{equation}\label{GT}
\mu=\mu_0+\Omega_v \kappa \gamma.
\end{equation}
Here $\Omega_v$ is the molecular volume and $\kappa$ is the mean curvature
of the surface (positive if the surface is convex).

Molecules will therefore have a tendency to leave convex portions of the
surface and attach to concave (or less convex) regions. In doing so, they 
may either travel through the bulk of the material outside the crystal, 
in a process of evaporation-condensation (or
solution-precipitation for a crystal against a melt), or they may diffuse
along the crystal surface \cite{Mullins}.

This leads to the crystal surface having a normal velocity $v_n$, which will
in general be the sum of a component $v_n^B$ from molecules arriving by 
diffusion through the bulk and $v_n^S$ from those arriving along the surface.
In sections \ref{BD} and \ref{SD} we consider limiting cases where 
only one of these mechanisms predominates. For vapor phase sintering, it is
likely that surface diffusion will dominate, while for liquid phase sintering
bulk diffusion may be the important mode \cite{Mullins}.
The case where crystal facets interfere \cite{Spohn} has been the 
subject of more recent work, but is not our concern in this paper.

Consider now two crystals which are close to their equilibrium shapes
(and therefore close to being spherical at least in the region of interest).
If these crystals are brought into gentle contact, the curvature near the
contact point will be large and negative, and therefore a neck will appear and
rapidly grow. There will usually be a grain boundary coincident with this 
neck, but in contrast to Ref. \cite{Mullins}, where equilibrium is approached
from an initial dihedral angle of $\pi$ radians, in the case of two sintering
crystals, the initial dihedral angle is $0$, and we no longer expect 
grain boundary grooving \cite{Mullins}.

The initial stages of neck growth occur rapidly, but nevertheless can be of 
importance in the rheology of crystal slurries, where contacts
may be formed and broken in rapid succession. Relevant examples might include
the flow of lava \cite{Pinkerton} or of brash ice \cite{Mellor}.

The initial growth of necks between two crystals has been analysed previously,
for example in Ref. \cite{Courtney} where a $\tilde{t}^{1/3}$ time dependence 
for neck radius as a function of time $\tilde{t}$ after contact was found 
for the case of bulk diffusion limited growth.

\begin{figure}
\includegraphics[width=3in]{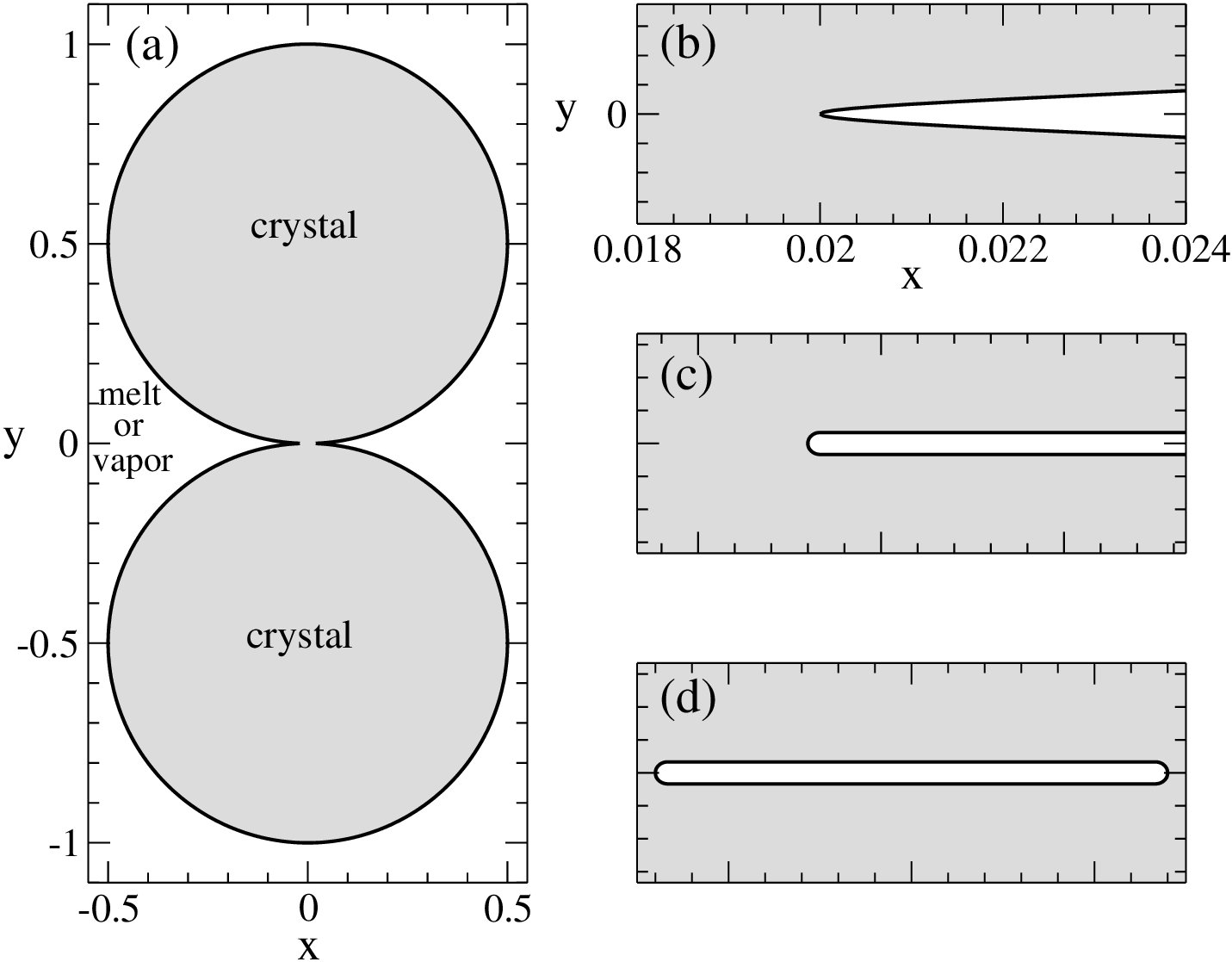}
\caption{\label{slot}
Image (a) shows schematically two spherical crystals that are in contact 
and have started to sinter together, forming a narrow neck of radius
$r_n=0.02$ (where all distances are scaled by a suitable capillary length). 
The crystals are axi-symmetric about an axis up the page.
In (b), a magnified portion of the neck region is depicted, illustrating that
at the very earliest stages of sintering, the surfaces of the two crystals are
nearly parallel close to the neck. Image (c) shows a 2 dimensional 
slot-shaped cavity, with exactly parallel surfaces, which we use as an
approximation to the axi-symmetric geometry in (b). Image (d) shows a
2d slot-shaped cavity of finite length, which is more appropriate for
simulations.}
\end{figure}

For the development of this paper, we note that there is a geometrical 
simplification which
occurs in the earliest stages of neck growth and which appears to have been
overlooked in the literature: If we consider two crystals just after they
have touched (Fig.~\ref{slot}) the parts of the surface near the contact
point (except for the very high curvature neck) are very nearly parallel 
to one another. This suggests that an interesting geometry for
investigating the early stages of neck growth is a $2$ dimensional
slot-shaped cavity, illustrated in Fig.~\ref{slot}(c). The closer in time
we are to first contact the better this is as an approximation. Furthermore,
if the radius of the neck is large compared to the separation of these
two nearly parallel surfaces (as is indeed the case at early times), then 
the difference between axi-symmetric cylindrical coordinates and 
plane Cartesian coordinates becomes negligible.

In sections \ref{BD} and \ref{SD} we analyse a slot-shaped cavity
of this kind, and then compensate for the remaining differences between
cylindrical and plane Cartesian geometry during structure evolution by
invoking conservation of volume. In this way, we obtain the geometry of
the crystals in the earliest moments after contact,
and also predictions for novel scaling behaviour of the neck size
as a function of time after contact.

\section{Bulk diffusion limited case}\label{BD}
Consider the case where growth is limited by bulk diffusion to and
from the crystal surfaces (attachment kinetics being unimportant
because the crystals are roughened).

We assume that the mean free path for bulk diffusion is
small compared to the feature sizes of interest. If $\phi({\rm\bf r})$
is the volume fraction in the vapor or melt of the type of molecules 
that can make up the crystal, then this quantity obeys a diffusion
equation with a diffusivity $D$.

We make the further approximation that the transients of this equation
have died out, so that the concentration field $\phi$ obeys Laplace's
equation
\begin{equation}\label{laplace}
\nabla^2\phi=0.
\end{equation}
This requires that the surface normal velocity $v_n$ should be small
compared to $D$ divided by a typical feature size.

The normal velocity $v_n$ of the crystal surface (which we denote by 
the set of points $S$) is obtained by conservation of volume for the
crystallizing material:
\begin{equation}\label{vn_BD}
v_n=v_n^B=D\left.\frac{\partial\phi({\rm\bf r})}{\partial {\rm\bf\hat{n}}}
\right|_{{\rm\bf r}\in S},
\end{equation}
where ${\rm\bf\hat{n}}$ is an outward unit normal vector to $S$.

We note in passing that we can apply the divergence theorem directly to
Eqs.~(\ref{vn_BD}) and (\ref{laplace}) with the result that the volume of 
an enclosed cavity is conserved under this evolution, while for
a discrete crystal in an open volume of vapour or melt, one must also
consider the concentration at infinity before it is possible to say
anything about conservation of crystal volume.

Let $\phi_0(T)$ be the concentration of molecules in the melt or
vapour which is present at equilibrium against a flat crystal surface
at temperature $T$, then
perturbing this linearly using Eq.~(\ref{GT}) we find that the actual 
concentration at the surface of a curved crystal is given by
\begin{equation}\label{phi_S}
\phi({\rm\bf r}\in S,T)=\phi_0(T)+\frac{\Omega_v T\gamma\kappa}{L}
\frac{d\phi_0}{dT},
\end{equation}
where $L$ is the latent heat per unit volume of fusion or evaporation.

Eqs.~(\ref{laplace},\ref{vn_BD},\ref{phi_S}) determine the evolution
of the system in an exactly analogous manner to the LSW
theory of Ostwald ripening \cite{Ostwald,Lifschitz,Wagner}.

Suppose the co-ordinates of a point in space are denoted by 
$(\tilde{x},\tilde{y},\tilde{z})$ and time after contact by $\tilde{t}$.
We define a capillary length $l_c\equiv\gamma\Omega_v/L$ and dimensionless
space, time and concentration variables $x=\tilde{x}/l_c$,
$t=\tilde{t}TD(d\phi_0/dT)/l_c^2$ and $\psi=(\phi-\phi_0)/(Td\phi_0/dT)$.

Eqs.~(\ref{laplace},\ref{vn_BD},\ref{phi_S}) then take the following simple
form in 2 dimensional Cartesian co-ordinates:
\begin{eqnarray}
\left(\frac{\partial^2}{\partial x^2}+\frac{\partial^2}{\partial y^2}
\right)\psi(x,y,t)=0 \label{BD1}\\
\psi\left[x,y_0(x,t),t\right]=
\frac{\partial^2 y_0/\partial x^2}{\left[
1+\left(\partial y_0/\partial x\right)^2\right]^{3/2}} \label{BD2}\\
\frac{\partial y_0}{\partial t}=-
\frac{\partial y_0}{\partial x}
\left.\frac{\partial\psi}{\partial x}\right|_{y=y_0}+
\left.\frac{\partial\psi}{\partial y}\right|_{y=y_0} \label{BD3}
\end{eqnarray}
where $y=y_0(x)$ is the crystal surface $S$.

\begin{figure}
\includegraphics[width=3in]{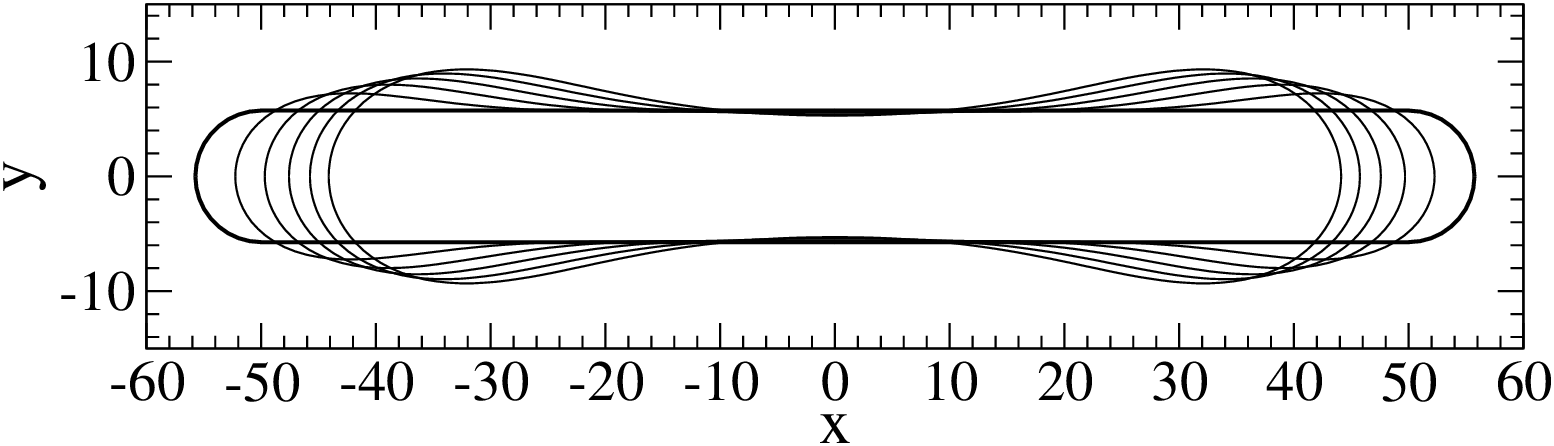}
\caption{\label{bulk_slot}
Numerical solution for the bulk diffusion limited evolution of a parallel-sided
2 dimensional slot-shaped cavity in a crystal. The initial surface is shown 
in bold, and subsequent points in time show the formation of dumbbell ends
to the curve. Although not shown, the final equilibrium shape must be a
circle at the origin containing the same area as the initial curve.}
\end{figure}

If we solve Eqs.~(\ref{BD1},\ref{BD2},\ref{BD3}) numerically, using a simple
explicit timestepping scheme for an initial slot-shaped cavity, we
see from Fig.~\ref{bulk_slot} that dumbbell-shaped ends start to appear
as the cavity evolves towards its eventual equilibrium shape, which must 
be a circle.

This behaviour suggests that there may be a travelling wave solution 
to Eqs.~(\ref{BD1},\ref{BD2},\ref{BD3}) which propagates at constant velocity
$v$ and without change of shape. 
This solution would consist of a teardrop-shaped
cavity formed from the evolution of a semi-infinite slot-shaped cavity in
which the parallel surfaces almost, but do not quite, touch. To picture this
slot, imagine the initial cavity of Fig.~\ref{bulk_slot}, but with two
changes: First, let it extend from $x=-\infty$ to $x=0$, rather than
the range $x=-56$ to $x=56$ (as shown in Fig.~\ref{bulk_slot}). Second, let 
it be very narrow, rather than extending between $y=-6$ and $y=6$
(as shown in Fig.~\ref{bulk_slot}).

If such a solution exists, then it is natural to tackle 
Eqs.~(\ref{BD1},\ref{BD2},\ref{BD3}) through complex analysis: Let
\begin{equation}
z\equiv x+iy,
\end{equation}
then if $\psi$ is an analytic function of $z$ (with an additional time 
dependence), Eq.~(\ref{BD1}) is satisfied as an identity, which simplifies the 
problem greatly.

\begin{figure}
\includegraphics[width=3in]{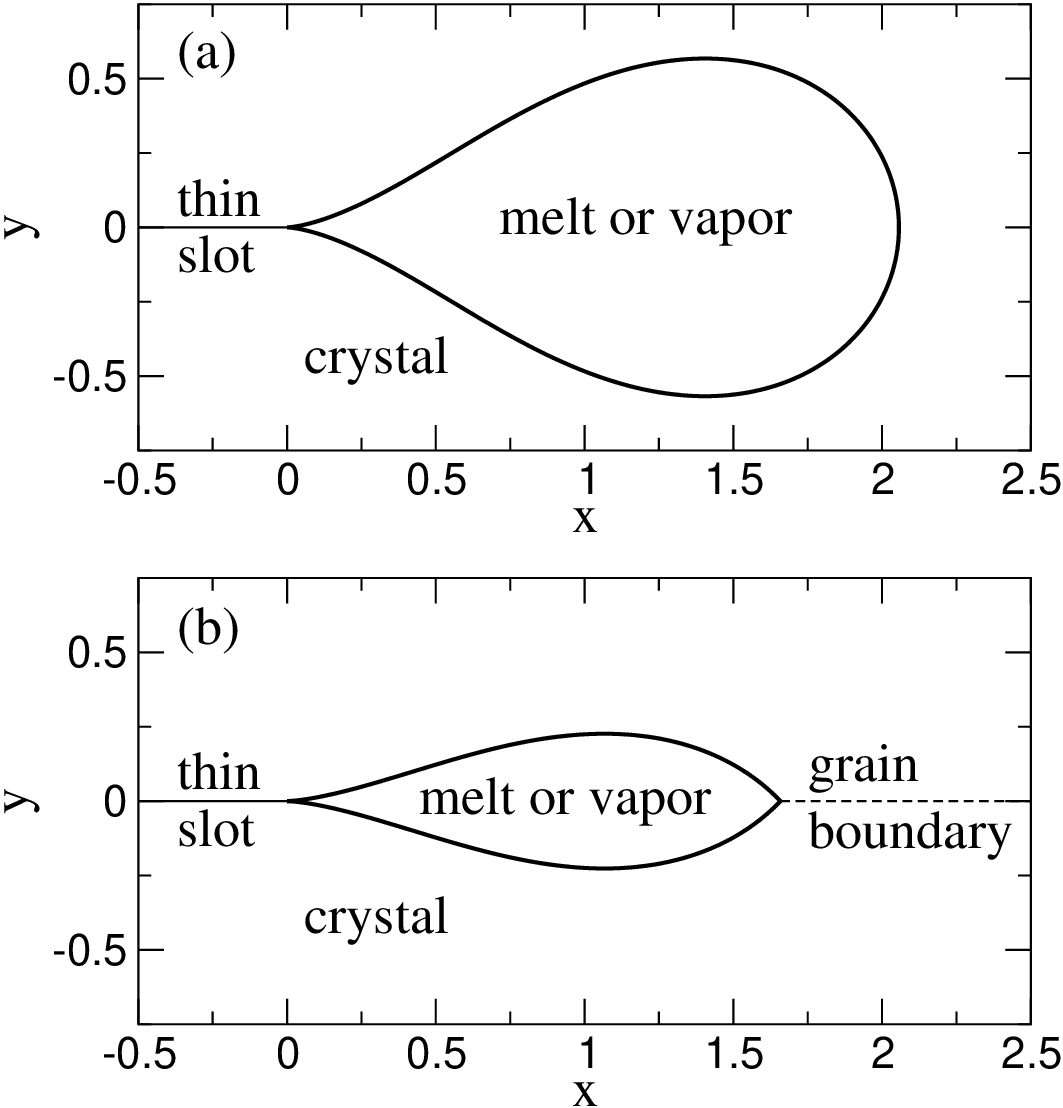}
\caption{\label{bulk_soliton}
Profiles of the travelling wave with $v=1$, for (a) $a=0.468703$ and (b)
$a=0.3$. The cavity moves to the left with speed $v=1$ without changing
shape. The thin (in this limit, zero thickness) parallel sided slot is
shown as the line from $x=0$ along the negative real axis
(which is the branch cut of Eq.~(\ref{seriespsi}). When $a\neq 0.468703$,
the profile has a cusp on the right hand side. This represents the (more
realistic) solution when the thin slot terminates at a grain boundary 
between two crystals, rather than being a cavity in a single crystal.}
\end{figure}

By balancing leading powers, we find that $\psi(z,t)$ must have a branch
point and the solution must be (up to translational symmetry) of the form
\begin{eqnarray}
\psi(z,t)\equiv v^{1/2}\hat{\psi}\left[ v^{1/2}(z+vt);a\right] \label{BSC1}\\
y_0(x,t)\equiv v^{1/2}\hat{y}_0\left[ v^{1/2}(x+vt);a\right], \label{BSC2}
\end{eqnarray}
where $a$ is a real parameter which represents a constant of integration in
Eqs.~(\ref{BD1},\ref{BD2},\ref{BD3}). Physically, $a$ determines the
curvature of the surface near to the tip of the teardrop-shaped cavity
(the origin in Fig.~\ref{bulk_soliton}).

After a considerable amount of back-substitution, one obtains
a series solution for the analytic function $\hat{\psi}(z;a)$ and
the real function $\hat{y}_0(x;a)$:

\begin{eqnarray}
\hat{\psi}(z;a)=az^{-1/2}-\frac{a^3}{18}z^{1/2}
-\frac{16}{15}z-\frac{5\ a^5}{1224}z^{3/2}-\frac{8\ a^2}{2835}z^2
\nonumber \\
-\left(\frac{7}{1950\ a}+\frac{1435\ a^7}{859248}\right)z^{5/2}
+\frac{3328\ a^4}{2284443}z^3+O(z^{7/2})
\label{seriespsi}
\\
\pm \hat{y}_0(x;a)=\frac{4\ a}{3}x^{3/2}+
\frac{38\ a^3}{27}x^{5/2}-\frac{8}{45}x^3+\frac{259\ a^5}{102}x^{7/2}
\nonumber \\
-\frac{1336\ a^2}{1701}x^4
+\left(\frac{10588243\ a^7}{1933308}-\frac{2}{8775\ a}\right)x^{9/2}
\nonumber \\
-\frac{95801752\ a^4}{34266645}x^5+O(x^{11/2}).
\label{seriesy0}
\end{eqnarray}

\begin{figure}
\includegraphics[width=3in]{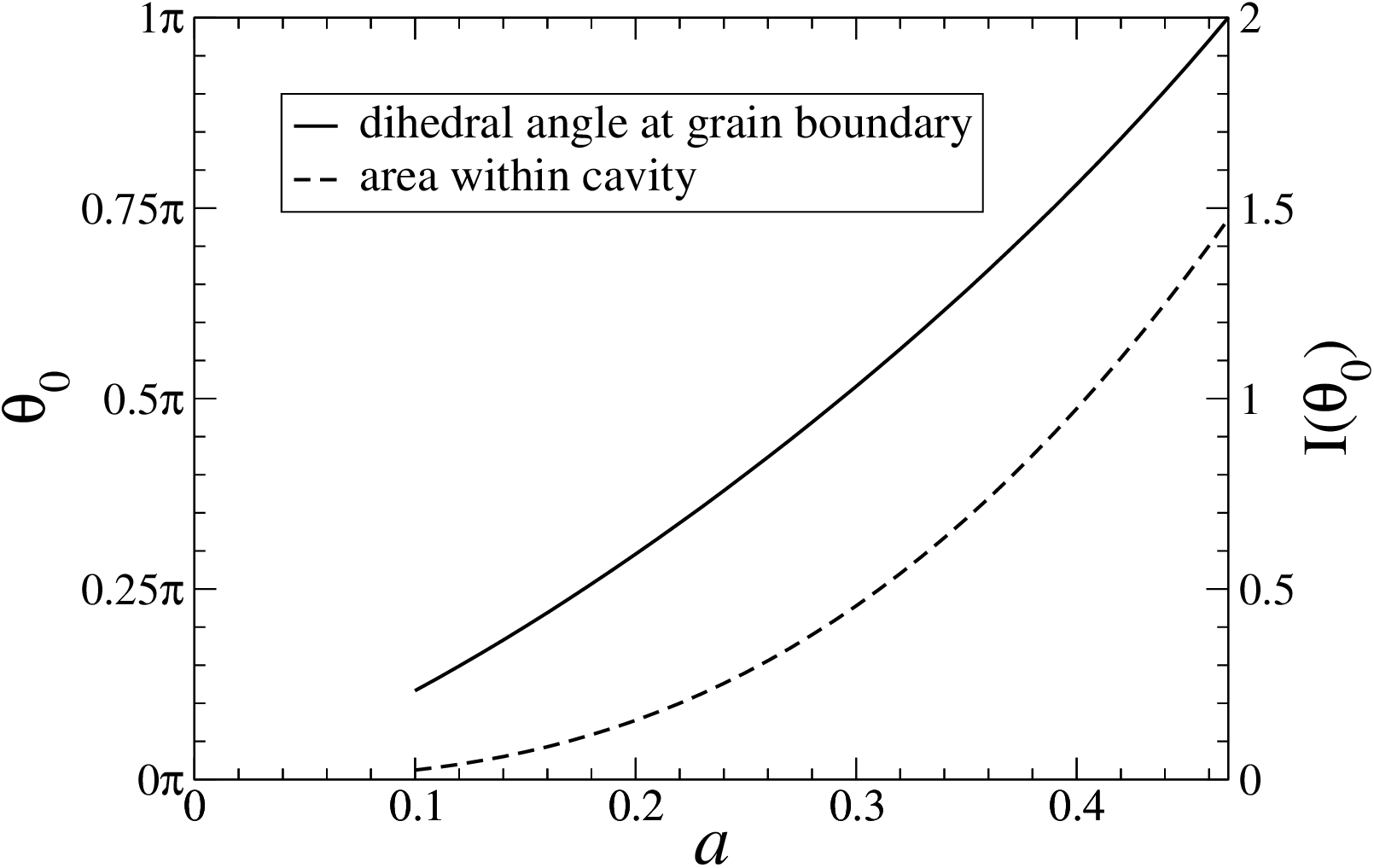}
\caption{\label{dihedral_angle}
Left hand scale and solid curve shows a plot of the dihedral angle 
$\theta_0$ at the grain boundary (if present) as a function of the
parameter $a$. The right hand scale and dashed curve shows the enclosed
area of the travelling wave cavity, defined by
$I(\theta_0)\equiv\int 2\hat{y}_0 \left[x;a(\theta_0)\right]dx$.
Both plots are for a (non-dimensional) wave velocity $v=1$.}
\end{figure}

The series solution for $\hat{y}_0(x)$ does not converge quickly enough
to be useful. However, Eq.~(\ref{seriespsi}) for $\hat{\psi}(z)$
does converge quickly in the region of interest, and using this solution,
the profile $\hat{y}_0(x)$ can be drawn numerically, using Eq.~(\ref{BD2}).
The results for two different values of the $a$ parameter are shown in
Fig.~\ref{bulk_soliton}, and a graph of the enclosed area in 
Fig.~\ref{dihedral_angle}.

\begin{figure}
\includegraphics[width=3in]{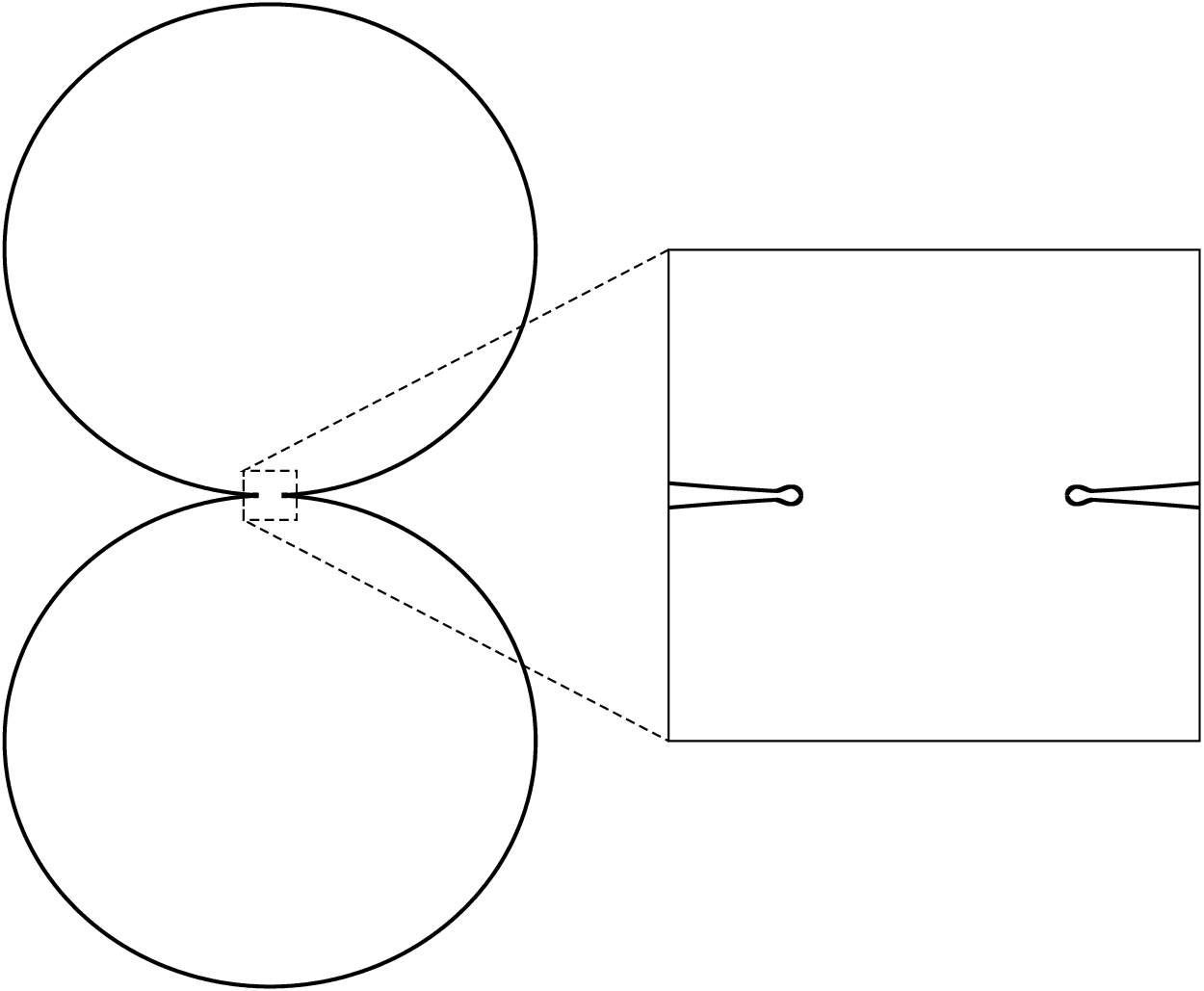}
\caption{\label{mag}
Schematic illustration of where the travelling wave solution of 
Fig.~\ref{bulk_soliton} is expected to occur during the sintering of two
spherical particles. Left hand illustration shows the two
crystals soon after contact. Right hand panel shows a magnified portion
of the neck region, with the teardrop travelling waves forming the
edges of the neck.}
\end{figure}

Although the travelling wave solution just presented is an idealized case,
an approximate analytical solution of this kind can also be used
to construct the rate of neck growth in the original problem of
two sintering spheres. This is done in the following manner:

From Eqs.~(\ref{BSC1},\ref{BSC2}) we see that under magnification the velocity
of the travelling wave is inversely proportional to its enclosed area,
and is independent of the separation of the parallel surfaces in the
initial slot-shaped cavity (provided this separation is very small). However,
if the separation between these surfaces is not zero, then the teardrop shaped 
cavity illustrated in Fig.~\ref{bulk_soliton} must grow in size as it
moves, in order to maintain conservation of enclosed area (or volume
in the three dimensional case), as discussed above. We note that for the
case of a neck growing between two touching spherical crystals,
the teardrop cavity which will form (shown schematically in Fig.~\ref{mag})
is not a completely enclosed volume, but will be nearly so at early times.

Now suppose we have two spherical crystals with initial (non-dimensionalized)
diameters $d_1$ and $d_2$, which are brought into contact. Let the
radius of the neck at (scaled) time $t$ after contact be $r_n(t)$.
The volume of material [to leading order in $r_n(t)/d_1$] that has been added 
to the crystals in order to make this solid neck is given by
\begin{equation}\label{neck_vol}
\pi (d_1+d_2)r_n^4(t)/(2d_1 d_2).
\end{equation}
If we assume that the neck region forms a nearly completely enclosed volume,
then the volume of crystal in this region must be conserved (using the
argument from the divergence theorem above). The material used to make the
solid neck must therefore come from the toroidal cavity formed by
the travelling wave (illustrated schematically in Fig.~\ref{mag}).

We can estimate this volume using Eq.~(\ref{BSC2}) and Pappus' second
centroid theorem \cite{Pappus} (that the volume generated by 
rotating a plane figure around
an axis is the product of its area and the distance moved by its centroid).
The result (again to leading order) is
\begin{equation}\label{torus_vol}
2\pi r_n(t) v^{-1}I(\theta_0),
\end{equation}
where
\begin{equation}
I(\theta_0)\equiv\int 2\hat{y}_{0}(x;a)dx
\end{equation}
is the enclosed area, which is a function of the equilibrium dihedral angle
$\theta_0$ at the grain boundary. The dihedral angle in turn is set 
by the grain boundary
surface energy per unit area $\gamma_{GB}$ through \cite{Mullins}
\begin{equation}
2\gamma\cos(\theta_0/2)=\gamma_{GB}.
\end{equation}

By equating the two volumes of Eqs.~(\ref{neck_vol}) and (\ref{torus_vol}), 
we obtain the final result for the growth velocity of the neck at
small times:
\begin{equation}\label{bulk_final}
\frac{dr_n(t)}{dt}\approx\frac{2}{r_n^3}I(\theta_0)
\frac{2d_1 d_2}{d_1+d_2},
\end{equation}
so that $r_n(t)\propto t^{1/4}$.

\section{Surface diffusion limited case}\label{SD}
If recrystallization is limited by the rate of surface diffusion instead
of bulk diffusion, then (still for the roughened, nearly isotropic
$\gamma$ case), the normal growth velocity of the crystal surface is
given by \cite{Mullins}
\begin{equation}\label{surf_ev}
v_n=v_n^S=\frac{\Omega_v^2 n_s D_s \gamma}{k_B T}\nabla_S^2 \kappa,
\end{equation}
where $n_s$ is the number of molecules per unit area at the surface 
available to diffuse, $D_s$ is a surface diffusivity, $k_B$ is
Boltzmann's constant and $\nabla_S^2$ is the surface Laplacian (also known
as the Laplace-Beltrami operator \cite{Beltrami}).

Under the action of Eq.~(\ref{surf_ev}), the volume of the crystal is
exactly conserved, which follows from the equivalent of the divergence
theorem for $\nabla_S^2$ \cite{Beltrami}.

In order to proceed, we non-dimensionalize in a similar manner to 
Sec.~\ref{BD}: Let the co-ordinates of a point in space be denoted by 
$(\tilde{x},\tilde{y},\tilde{z})$ and time after contact by $\tilde{t}$.
We define a capillary length $l_{s}\equiv\gamma\Omega_v/k_B T$ and 
dimensionless space and time variables $x=\tilde{x}/l_{s}$ and
$t=\tilde{t}\Omega_v n_s D_s/l_{s}^3$.

The equation for surface evolution then takes the non-dimensionalized
form
\begin{equation}\label{surf_ev_simp}
v_n=\nabla_S^2\kappa.
\end{equation}

\begin{figure}
\includegraphics[width=3in]{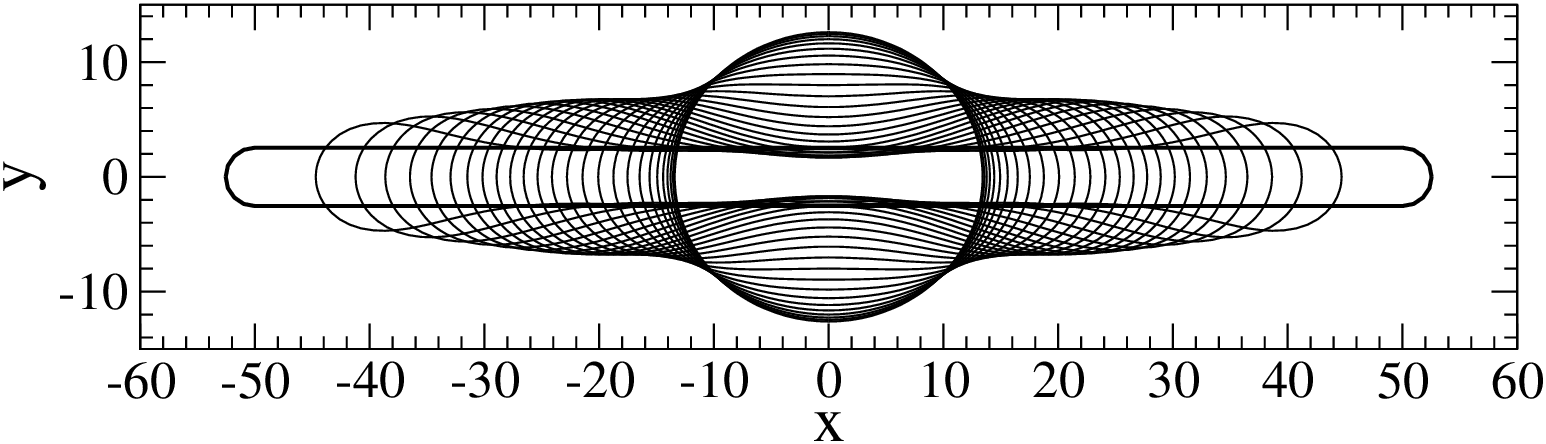}
\caption{\label{surface_slot}
Numerical solution for the surface diffusion limited evolution of a thin,
parallel-sided slot-shaped cavity in a crystal. The initial surface is
shown in bold, and subsequent moments in time show the formation of
dumbbell-shaped ends to the curve, before the final equilibrium shape
is reached, which is a circle enclosing the same area as the original
curve.}
\end{figure}

If we again perform a simple, explicit numerical simulation for the
evolution of a slot-shaped cavity in 2 dimensions, we obtain the 
results shown in Fig.~\ref{surface_slot}.

This problem has in fact been studied before \cite{Lee,Kampe}, in the context 
of the `inverse' case, where
the authors study the evolution and breakup of a thin plate. However,
since the evolution is determined by the surface only, this does not affect the
equations. The formation of dumbbell shaped ends in Fig.~\ref{surface_slot}
is then the analogue of the first stages of edge spheroidization of a plate
\cite{Lee,Nichols}.

Once more, this figure suggests that it is worth seeking a travelling
wave solution to the evolution equation, which moves at constant
velocity $v$ without change of shape. To do this, we note that if
$\Theta$ is the angle that the curve representing the crystal surface
makes with the $y$-axis, $s$ is
the dimensionless distance along this curve and $v$ is the dimensionless
speed of propagation, then the relevant equation to solve is
[from Eq.~(\ref{surf_ev_simp})]
\begin{equation}\label{surf_sol}
v\cos\Theta=\frac{d^3\Theta(s)}{ds^3}
\end{equation}
with boundary condition $\Theta(0)=d^2\Theta(0)/ds^2=0$.

\begin{figure}
\includegraphics[width=3in]{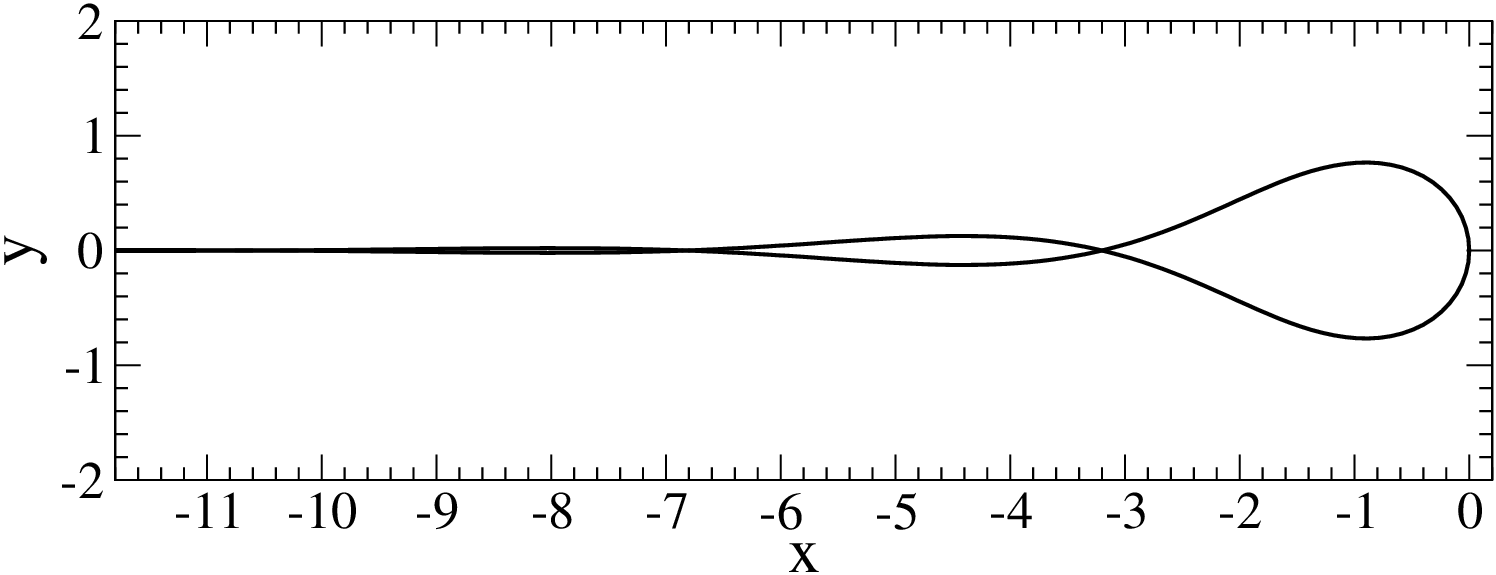}
\caption{\label{surface_soliton}
Travelling wave solution for the equation of surface diffusion limited
evolution [Eq.~(\ref{surf_ev_simp})]. The entire curve moves to the
left at speed $v=1$. This is a purely mathematical solution to the problem,
as the curve intersects itself and so cannot represent a real crystal
boundary. The total (algebraic and non-dimensionalized) area enclosed 
by the curve is 2.536}
\end{figure}

Fig.~\ref{surface_soliton} shows a numerical solution corresponding to
$v=1$, while the form of Eq.~(\ref{surf_sol}) shows that under magnification,
the velocity $v$ is proportional to the enclosed area to the power $-3/2$.
Although a formal solution to Eq.~(\ref{surf_sol}), the curve is
self-intersecting; indeed the assymptotic form for large negative $x$ is
\begin{equation}
y_0(x)\sim \pm e^{x/2}\cos(x\sqrt{3}/2).
\end{equation}

\begin{figure}
\includegraphics[width=3in]{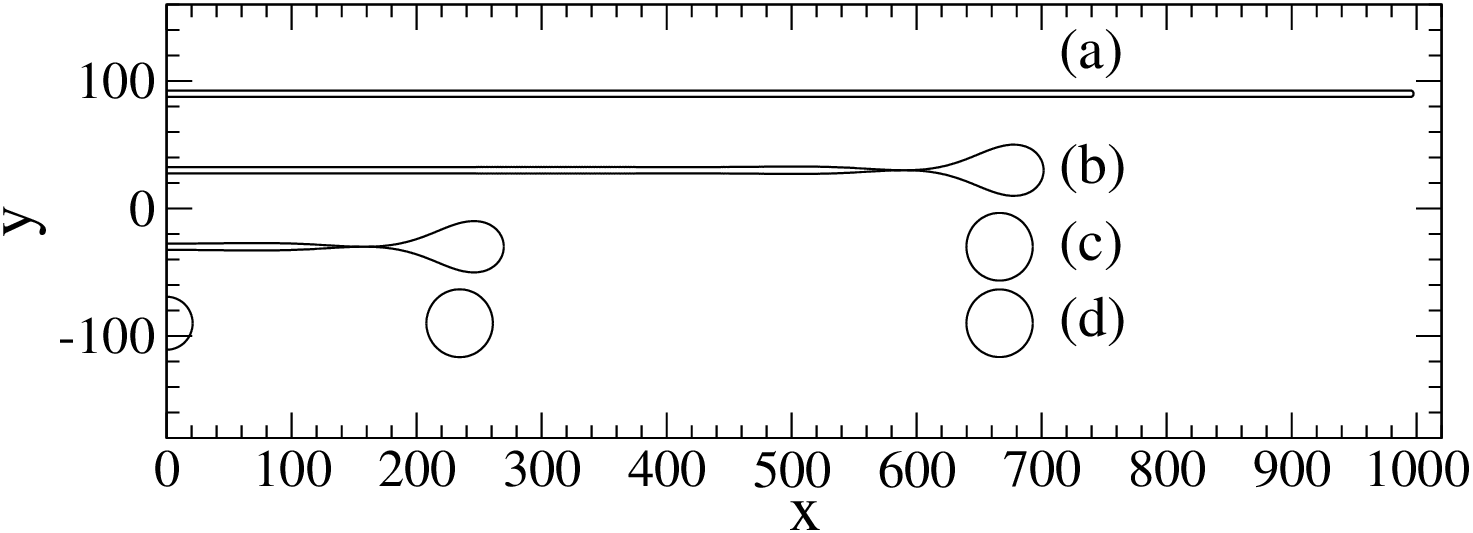}
\caption{\label{surface_long_slot}
Snapshots in time (displayed with vertical offset for clarity) for the
evolution of an initial long slot-shaped cavity under
surface diffusion limited evolution [Eq.~(\ref{surf_ev_simp})].
The initial cavity is shown in (a) and has a dimensionless width $w=16/\pi$.
Configuration (b) is the first pinch-off event, at dimensionless
time $t=2.17\times 10^6$. Image (c) shows the second pinch-off event
at $t=4.34\times 10^6$ and (d) is the final, equilibrium configuration.
Curves are drawn only for $x>0$, and are in fact symmetric under
reflection about $x=0$.}
\end{figure}

This self-intersecting shape strongly suggests that under the evolution
equation [Eq.~(\ref{surf_ev_simp})], a long slot-shaped cavity will
repeatedly pinch off to form a string of equally spaced circular cavities
in its wake. This behaviour is indeed seen in the numerical
simulation of a very long cavity in Fig.~\ref{surface_long_slot}.

From the simulation results of Fig~\ref{surface_long_slot} and the scaling
properties of Eq.~(\ref{surf_sol}), we see that for a long slot-shaped
cavity with a dimensionless separation $w$ between the opposite,
parallel crystal surfaces, the average (dimensionless) velocity of the end 
of the slot will be
\begin{equation}\label{slot_velocity}
\langle v\rangle\approx 0.0263 w^{-3},
\end{equation}
and the radius of each circle left behind in the wake is approximately
$5.21w$.

For a pair of spherical crystals just after contact and evolving according to
Eq.~(\ref{surf_ev_simp}), it would be tempting to use Eq.~(\ref{slot_velocity}),
replacing $w$ with the separation of the undisturbed spherical surfaces.
This would give an average rate of growth of the radius $r_n$ of the
neck proportional to $r_n^{-6}$. However, this 
would be incorrect, as conservation
of crystal volume is not taken into account properly.

Instead, for the axi-symmetric case of two spherical crystals each of 
(dimensionless) diameter $d$, we impose conservation of crystal volume
in the following manner: First, assume that pinching-off produces a
series of concentric tori with major radii given by the set
$\{ r_i \}$ with $i\in \{1,2 \ldots\}$, and minor radii by the set
$\{\rho_i\}$ given by
\begin{equation}
\rho_i\approx 5.21 w(r_i)\equiv 5.21\left( d-\sqrt{d^2-4r_i^2}\right)
\end{equation}
($w(r)$ being the separation of the surfaces of the original spheres
at distance $r$ radially from the initial contact point).

Conservation of enclosed volume then allows us to write down the distance
from the axis of symmetry at which the $i$'th torus lies \cite{Pappus}:
\begin{equation}\label{tori}
\int_{r_{i-1}}^{r_i}2\pi r w(r)dr
\approx 2\pi r_i\pi\left[ 5.21 w(r_i)\right]^2.
\end{equation}
However, Eq.~(\ref{tori}) has only one solution, namely
\begin{equation}\label{one_torus}
r_1\approx\frac{d}{8\pi (5.21)^2}\approx 0.00147 d.
\end{equation}
We therefore expect the growing neck to leave behind one tiny toroidal
cavity, before proceeding to grow without pinch-off events. 
We conjecture that this toroidal
cavity will subsequently break up into a ring of spheres by a similar
mechanism to the breakup of a cylinder \cite{Nichols} or edge spheroidization
of a plate \cite{Lee}.

If, after this single pinch-off event, the profile still maintains a 
dumbbell-shaped end, similar to Fig.~\ref{surface_long_slot}, and containing
most of the volume of the vapor or melt in the neck region, then
(again by conservation of enclosed volume and using the algebraic
area in the curve of Fig.~\ref{surface_soliton}), the neck radius $r_n$
should grow as
\begin{equation}\label{surface_final}
\frac{dr_n(t)}{dt}\approx\left(
\frac{2\times 2.536 r_n d}{r_n^4-(0.00147d)^4}
\right)^{3/2},
\end{equation}
so that very approximately, $r_n(t)\propto t^{1/3}$.

We note that grain boundaries are also easily incorporated into this formalism,
by changing the value of $\Theta(0)$ in Eq.~(\ref{surf_sol}) to introduce the
relevant dihedral angle at the origin in Fig.~\ref{surface_soliton}.
Just as in section \ref{BD}, the grain boundary runs parallel to the
slot (and outside it), while in Ref.~\cite{Lee} the geometry is very different:
grain boundaries run perpendicular to the plate (and inside it).

\section{Conclusions}
For two sintering roughened crystals, the observation that the geometry near 
the contact point in the first 
moments after contact is two dimensional (and very similar to a
slot-shaped cavity) is a considerable simplification for analysing the problem
at these early times.

We are able to bring to bear complex analysis in the case of diffusion
limited crystal evolution, and simple numerical approaches for the
surface diffusion limited case.

The result is a novel scaling for the power law for growth of the neck
in the diffusion limited case [namely
$r_n(t)\propto t^{1/4}$ from Eq.~(\ref{bulk_final})], and the prediction of
a pinch-off event [Eq.~(\ref{one_torus})] followed by approximate
power law growth [Eq.~(\ref{surface_final})] for the surface diffusion
limited case.

We also believe that the travelling wave solutions pivotal to these
results (namely Figs.~\ref{bulk_soliton} and \ref{surface_soliton}) have 
some aesthetic appeal.

\acknowledgments
The authors thank Ian Burns, Michael van Ginkel, Scott Singleton, Chris Clarke,
Javier Aldazabal and Aitor Luque for many useful discussions.

All figures were prepared using `Grace' (http://plasma-gate.weizmanm.ac.il),
while Eqs.~(\ref{seriespsi}) and (\ref{seriesy0}) were derived with the aid
of `Mathematica' (http://www.wolfram.com).

\end{document}